\documentclass{epl}

\usepackage{graphicx}
\usepackage{amssymb}

\newcommand{\bra}[1]{\langle #1|}
\newcommand{\ket}[1]{|#1\rangle}

\title{Cotunneling through two-level quantum dots weakly coupled to ferromagnetic leads}
\shorttitle{Cotunneling through two-level quantum dots}

\author{Ireneusz Weymann}

\institute{Department of Physics, Adam Mickiewicz University,
61-614 Pozna\'n, Poland}


\begin{document}

\maketitle

\begin{abstract}
The spin-polarized transport through two-level quantum dots weakly
coupled to ferromagnetic leads is considered theoretically in the
Coulomb blockade regime. It is assumed that the dot is doubly
occupied, so that the current flows due to cotunneling through
singlet and triplet states of the dot. It is shown that transport
characteristics strongly depend on the ground state of quantum
dot. If the ground state is a singlet, differential conductance
($G$) displays a broad minimum at low bias voltage, while tunnel
magnetoresistance (TMR) is given by the Julliere value. If triplet
is the ground state of the system, there is a maximum in
differential conductance at zero bias when the leads are
magnetized in antiparallel. The maximum is accompanied by a
minimum in TMR. The different behavior of $G$ and TMR may thus
help to determine the ground state of the dot and the energy
difference between the singlet and triplet states.
\end{abstract}

\section{Introduction}

Transport properties of quantum dots coupled to ferromagnetic
leads have been a subject of thorough studies since a few years
\cite{loss02,maekawa02}. The considerations concerned both the
strong coupling regime where the Kondo physics emerges
\cite{kondo}, as well as the weak coupling regime. In the weak
coupling regime most of theoretical works addressed the problem of
spin-dependent sequential transport through quantum dots hosting
single orbital level \cite{sequential}. In the sequential
tunneling regime, if the applied bias voltage exceeds a certain
threshold voltage, electrons tunnel one by one through the system,
otherwise the current is suppressed leading to the Coulomb
blockade effect. Although in the Coulomb blockade regime the
sequential transport is suppressed, the current can be still
mediated by higher-order tunneling processes such as cotunneling
which involves correlated tunneling of two electrons {\it via}
virtual states of the dot \cite{cotunneling}. Spin-polarized
cotunneling transport has been addressed very recently. It has
been shown that when the dot is singly occupied, a zero-bias
anomaly appears in differential conductance when the
magnetizations of the leads are aligned in antiparallel
\cite{weymanndong,weymannPRB}. These considerations were performed
only for single-level quantum dots. In real systems, however,
usually more than one energy level participate in transport, which
can lead to further interesting effects \cite{twolevelQD}.

The goal of this paper is to address the problem of spin-dependent
transport through two-level quantum dots coupled to ferromagnetic
leads in the cotunneling regime. More specifically, we analyze the
case when the dot is doubly occupied at equilibrium and the system
is in the Coulomb blockade regime. In this case, depending on the
value of the exchange interaction, either the singlet or triplet
state will be favored. We show that the differential conductance
($G$) and tunnel magnetoresistance (TMR) display a strong
dependence on the ground state of the system. If the ground state
is a singlet, $G$ exhibits a broad minimum at low bias voltage in
both magnetic configurations of the system, while TMR takes the
Julliere value \cite{julliere75}. On the other hand, if the ground
state is a triplet, there is a maximum in differential conductance
at zero bias in the antiparallel configuration, accompanied by a
minimum in TMR. The different behavior of both differential
conductance and TMR provides thus a handle to determine the ground
state of the system.

The systems discussed in this paper may be realized experimentally
for example in metallic single-wall carbon nanotube (SWCNT)
quantum dots contacted to ferromagnetic leads \cite{FMSWCNT}. As
shown recently, the shell filling of SWCNT quantum dots exhibits
four-electron periodicity when sweeping the gate voltage
\cite{SWCNT}. By changing the gate, one can tune the system to the
Coulomb blockade regime where the dot is doubly occupied at
equilibrium.

\section{Model and method}

\begin{figure}[b]
  \center
  \includegraphics[width=0.4\columnwidth]{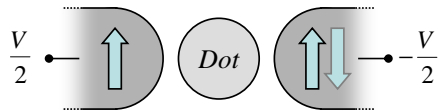}
  \caption{\label{Fig:1} Schematic
  of a quantum dot coupled to ferromagnetic leads.
  The magnetizations of the leads can form either
  parallel or antiparallel magnetic configuration.
  The system is symmetrically biased.}
\end{figure}

We consider spin-polarized cotunneling transport through quantum
dots with two orbital levels coupled to ferromagnetic leads. It is
assumed that the magnetic moments of the leads can form either
parallel or antiparallel magnetic configuration, see
fig.~\ref{Fig:1}. The Hamiltonian $H$ of the system involves four
terms, $H=H_{\rm L} + H_{\rm R} + H_{\rm D} + H_{\rm T}$. The
first two terms describe the noninteracting itinerant electrons in
the leads,
$H_r=\sum_{k\sigma}\varepsilon_{rk\sigma}c^{\dagger}_{rk\sigma}c_{rk\sigma}$,
for $r={\rm L,R}$, corresponding to the left and right leads, with
$\varepsilon_{rk\sigma}$ being the energy of an electron with wave
number $k$ and spin $\sigma$ in the lead $r$, and
$c^{\dagger}_{rk\sigma}$ ($c_{rk\sigma}$) denoting the respective
creation (annihilation) operator. The quantum dot is described by
the Hamiltonian \cite{hamiltonian}
\begin{equation}
  H_{D} =\sum_{j\sigma} \varepsilon_{j} n_{j\sigma}
  + U \sum_j n_{j\uparrow} n_{j\downarrow}
  + U^\prime \sum_{\sigma\sigma^\prime} n_{1\sigma}n_{2\sigma^\prime}
  - J \sum_{\alpha\beta\gamma\delta} d^\dagger_{1\alpha}
    d_{1\beta} d^\dagger_{2\gamma} d_{2\delta}
    \vec{\sigma}_{\alpha\beta} \vec{\sigma}_{\gamma\delta} \,,
\end{equation}
where $n_{j\sigma}=d^{\dagger}_{j\sigma}d_{j\sigma}$ and
$d^{\dagger}_{j\sigma}$ ($d_{j\sigma}$) is the creation
(annihilation) operator of an electron with spin $\sigma$ on the
$j$th level ($j=1,2$), while $\varepsilon_{j}$ is the
corresponding energy. On-level and inter-level Coulomb interaction
between electrons is described by $U$ and $U^\prime$,
respectively. The last term in $H_D$ corresponds to the exchange
energy due to the Hund's rule, with $J$ being the respective
exchange coupling and $\vec{\sigma}$ denoting a vector of Pauli
spin matrices. To simplify further discussion of numerical results
we assume $U^\prime=U$. For the energy of the first orbital level
we write $\varepsilon_1 = \varepsilon$ and $\varepsilon_2 =
\varepsilon + \delta\varepsilon$, where $\delta\varepsilon$ is the
splitting of the dot orbital levels. For example in metallic SWCNT
quantum dots, $\delta\varepsilon$ would correspond to the energy
mismatch between the two subbands of the nanotube \cite{SWCNT}.
The mismatch can vary with the boundary conditions at the
nanotube-electrode interface.

The last term of $H$ describes the tunneling processes between the
dot and electrodes
\begin{equation}
  H_{\rm T}=\sum_{r=\rm
  L,R}\sum_{jk\sigma}\left(T_{rj}c^{\dagger}_{r k\sigma}
  d_{j\sigma}+ T_{rj}^\star d^\dagger_{j\sigma} c_{r k\sigma}
  \right) \,,
\end{equation}
with $T_{rj}$ denoting the tunnel matrix elements between the lead
$r$ and $j$th level. Coupling of the $j$th dot level to external
leads can be described by $\Gamma_{rj}^{\sigma}= 2\pi |T_{rj}|^2
\rho_r^\sigma$, with $\rho_r^\sigma$ being the spin-dependent
density of states in the lead $r$. By introducing the spin
polarization of lead $r$, $p_{r}=(\rho_{r}^{+}- \rho_{r}^{-})/
(\rho_{r}^{+}+ \rho_{r}^{-})$, the couplings can be expressed as,
$\Gamma_{rj}^{+(-)}=\Gamma_{rj}(1\pm p_{r})$, with $\Gamma_{rj}=
(\Gamma_{rj}^{+} +\Gamma_{rj}^{-})/2$. Here, $\Gamma_{rj}^{+}$ and
$\Gamma_{rj}^{-}$ describe the coupling of the $j$th level to
spin-majority and spin-minority electron bands, respectively. We
assume that the system is coupled symmetrically to the leads
$\Gamma_{rj}\equiv\Gamma/2$ and $p_{\rm L}=p_{\rm R}\equiv p$. As
reported in ref.~\cite{kogan04}, typical values of $\Gamma$ in the
weak coupling regime are of the order of tens of $\mu$eV.

In order to find the current flowing through the dot in the
Coulomb blockade, one has to calculate the respective cotunneling
rates. Within the second-order perturbation theory, the rate of a
cotunneling process from lead $r$ to lead $r^\prime$ which changes
the dot state from $\ket{\chi}$ into $\ket{\chi^\prime}$ is given
by
\begin{equation}\label{Eq:cotunnelingrate}
  \gamma_{\rm rr^\prime}^{\chi\rightarrow\chi^\prime} = \frac{2\pi}{\hbar}
  \left|\sum_{v}\frac{\bra{\Phi_{r^\prime}^{\chi^\prime}}H_{\rm T}\ket{\Phi_v}\bra{\Phi_v}
  H_{\rm T} \ket{\Phi_{r}^\chi}} {\varepsilon_i-\varepsilon_{v}}\right|^2\delta
  (\varepsilon_i-\varepsilon_f),
\end{equation}
with $\varepsilon_i$ and $\varepsilon_f$ denoting the energies of
initial and final states, $\ket{\Phi_r^\chi}$ being the state of
the system with an electron in the lead $r$ and the dot in state
$\ket{\chi}$, whereas $\ket{\Phi_v}$ is a virtual state with
$\varepsilon_{v}$ denoting the corresponding energy. Among
different cotunneling processes one can distinguish the
single-barrier ($r=r^\prime$) and double-barrier ($r\neq
r^\prime$) cotunneling as well as spin-flip ($\chi\neq
\chi^\prime$) and non-spin-flip ($\chi= \chi^\prime$) cotunneling.
The spin-flip processes change the state of the dot, whereas the
non-spin-flip processes are fully coherent, i.e. they do not
change the dot state. The current flows through the system due to
double-barrier cotunneling processes. On the other hand, the
single-barrier processes do not contribute directly to electric
current, however, they can change the dot occupations, and this
way also the current. The details of how to determine different
cotunneling rates can be found in ref. \cite{cotunneling}.

The cotunneling current flowing through the system from the left
to right lead is given by
\begin{equation}
  I = e \sum_{\chi\chi^\prime} P_\chi \left[
  \gamma^{\chi \rightarrow \chi^\prime}_{\rm LR} -
  \gamma^{\chi \rightarrow \chi^\prime}_{\rm RL} \right],
\end{equation}
where $P_\chi$ denotes the corresponding occupation probability.
The probabilities $P_\chi$ can be found from the master equation,
$0= \sum_{rr^\prime}\sum_{\chi^\prime} \left[-\gamma_{rr^\prime}
^{\chi\rightarrow\chi^\prime} P_{\chi} + \gamma_{rr^\prime}
^{\chi^\prime\rightarrow\chi} P_{\chi^\prime}\right]$, together
with the normalization condition $\sum_{\chi}P_\chi = 1$.

We consider the case when the dot is doubly occupied at
equilibrium and the system is in the Coulomb blockade regime.
There are then six different two-particle states in the dot
possible, these are three singlets $\ket{S=0,M=0}_1 =
(\ket{\uparrow}\ket{\downarrow} -
\ket{\downarrow}\ket{\uparrow})/\sqrt{2}$, $\ket{0,0}_2 =
\ket{\uparrow\downarrow} \ket{0}$, $\ket{0,0}_3 = \ket{0}
\ket{\uparrow\downarrow}$, and three triplets
$\ket{1,0}=(\ket{\uparrow}\ket{\downarrow} +
\ket{\downarrow}\ket{\uparrow})/\sqrt{2}$,
$\ket{1,1}=\ket{\uparrow}\ket{\uparrow}$ and
$\ket{1,-1}=\ket{\downarrow}\ket{\downarrow}$, where the first
(second) {\it ket} corresponds to the first (second) orbital level
of the dot. In the case of finite level spacing,
$\delta\varepsilon>k_{\rm B}T,\Gamma$, and $J<\delta\varepsilon$,
the lowest singlet state is $\ket{0,0}_2$. In the following, we
show that transport characteristics strongly depend on the ground
state of the system. It is therefore useful to introduce the
difference between the energy of the lowest lying singlet
$(\varepsilon_{\rm S})$ and triplet $(\varepsilon_{\rm T})$
states, $\Delta_{\rm ST} = \varepsilon_{\rm S} - \varepsilon_{\rm
T} = J - \delta\varepsilon$.

\section{Results and discussion}

\begin{figure}[t]
  \center
  \includegraphics[width=0.75\columnwidth]{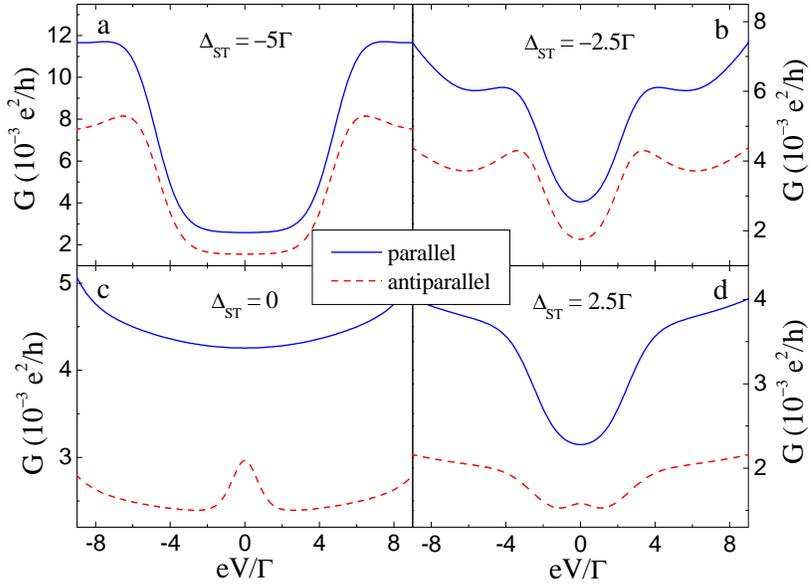}
  \caption{\label{Fig:2} (color online) The differential conductance
  in the parallel (solid line) and antiparallel (dashed line)
  magnetic configuration of the system
  as a function of the bias voltage for several values of
  $\Delta_{\rm ST}= J-\delta\varepsilon$
  as indicated: (a) $\Delta_{\rm ST}=-5\Gamma$,
  (b) $\Delta_{\rm ST}=-2.5\Gamma$, (c) $\Delta_{\rm ST}=0$,
  (d) $\Delta_{\rm ST}=2.5\Gamma$.
  The other parameters are: $k_{\rm B}T=0.5\Gamma$,
  $\varepsilon=-60\Gamma$, $\delta\varepsilon = 5\Gamma$,
  $U = U^\prime = 40\Gamma$, and $p=0.5$.}
\end{figure}

Figure~\ref{Fig:2} presents the bias voltage dependence of
differential conductance $G$ in the parallel and antiparallel
magnetic configuration for several values of $\Delta_{\rm ST} = J-
\delta\varepsilon$. First of all, one can see that the transport
characteristics show a distinctively different behavior depending
on $\Delta_{\rm ST}$, i.e. the ground state of the quantum dot.
For $\Delta_{\rm ST}<0$, the ground state of the dot is a singlet,
$\ket{0,0}_2$,  whereas for $\Delta_{\rm ST}>0$, the ground state
is a triplet, which is three-fold degenerate, $\ket{1,0}$,
$\ket{1,1}$, $\ket{1,-1}$. On the other hand, for $\Delta_{\rm
ST}=0$, the dot is in a mixed state and the occupation of singlet
and each triplet is equal at equilibrium and given by $1/4$.

First, let us discuss the case of $\Delta_{\rm ST}<0$ shown in
fig.~\ref{Fig:2}(a) and (b). In this situation, at low bias
voltage the dot is occupied by two electrons on the lowest energy
level, $\ket{0,0}_2 = \ket{\uparrow\downarrow}\ket{0}$, and the
ground state is singlet, $S=0$. Because for $\Delta_{\rm ST}<0$
and $|eV|<|\Delta_{\rm ST}|$, $P_{\ket{0,0}_2}\approx 1$, the
current is mediated only by non-spin-flip cotunneling processes.
However, once $|eV|\gtrsim |\Delta_{\rm ST}|$, the triplet states
start participating in transport (in the case of $\Delta_{\rm
ST}=-5\Gamma$ also the state $\ket{0,0}_1$ contributes), leading
to an increase in the differential conductance at
$|eV|\approx|\Delta_{\rm ST}|$. In other words, the suppression of
cotunneling through $S=1$ states gives rise to a broad minimum in
differential conductance which is present in both magnetic
configurations of the system, see fig.~\ref{Fig:2}(a) and (b). The
width of this minimum is determined by the splitting between the
singlet and triplet states, $2|\Delta_{\rm ST}|$. By measuring the
width, one can thus obtain information about the energy difference
between the $S=0$ and $S=1$ states.

In the case when the current flows only due to non-spin-flip
cotunneling through the state $\ket{0,0}_2$, one can find
approximative formulas for the differential conductance. At low
temperature and for $\delta\varepsilon \gg k_{\rm B}T$, $G$ in the
parallel configuration can be expressed as
\begin{equation}
  G_{\rm P}^{S=0} =
  \frac{e^2\Gamma^2}{2h} (1+p^2)\left[
  \frac{1}{(\varepsilon+U)^2}
  + \frac{1}{(\varepsilon+\delta\varepsilon+2U)^2}
  \right] \,,
\end{equation}
on the other hand, for the antiparallel configuration one finds,
$G_{\rm AP}^{S=0} = (1-p^2)G_{\rm P}^{S=0}/(1+p^2)$. These
expressions approximate the minimum in differential conductance at
low bias voltage shown in fig.~\ref{Fig:2}(a). The analytical
formulas for the general case when more states participate in
transport are too cumbersome to be presented here.

\begin{figure}[h]
  \center
  \includegraphics[width=0.42\columnwidth]{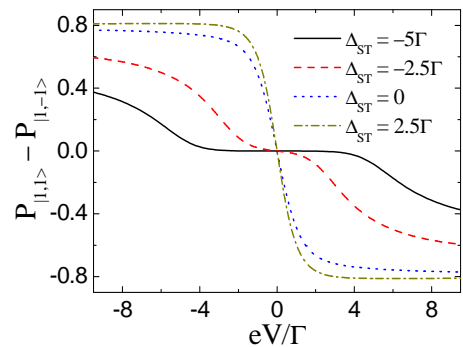}
  \caption{\label{Fig:3} (color online)
  Nonequilibrium spin accumulation in the dot,
  $P_{\ket{1,1}}-P_{\ket{1,-1}}$,
  as a function of the bias voltage for
  $\Delta_{\rm ST} = -5,-2.5,0,2.5 \Gamma$.
  The other parameters are the same as in fig.~\ref{Fig:2}.}
\end{figure}

Figure~\ref{Fig:2}(c) displays the differential conductance in the
case when the singlet and triplet states are degenerate,
$\Delta_{\rm ST}=0$. First of all, it can be seen that $G$
exhibits a distinctively different behavior in both magnetic
configurations. In the parallel configuration the differential
conductance has a smooth parabolic dependence on the bias voltage,
whereas in the antiparallel configuration there is a maximum at
the zero bias voltage. This effect bears a resemblance to that
found in the case of singly occupied one-level quantum dots
\cite{weymanndong,weymannPRB}. Here, however, the mechanism
leading to the maximum is different -- the zero-bias peak appears
due to cotunneling through singlet and triplet states of the dot.
In the case of $\Delta_{\rm ST}=0$ and at low bias voltage, all
the four dot states, i.e. $\ket{0,0}_2$, $\ket{1,0}$,
$\ket{1,-1}$, $\ket{1,1}$, participate in transport on an equal
footing. Consequently, the current flows due to both spin-flip and
non-spin-flip cotunneling processes. To understand the mechanism
leading to the zero-bias peak, one should bear in mind that in the
antiparallel configuration the spin-majority electrons of one lead
tunnel to the spin-minority electron band of the other lead. For
example, for positive bias voltage (electrons tunnel then from the
right to left lead), the spin-$\uparrow$ electrons can easily
tunnel to the left lead (the spin-$\uparrow$ electrons are the
majority ones), while this is more difficult for the
spin-$\downarrow$ electrons (they tunnel to the minority electron
band). Thus, with increasing the bias voltage, the occupation of
state $\ket{1,-1}$ ($\ket{\downarrow}\ket{\downarrow}$) is
increased, while the occupation of state $\ket{1,1}$
($\ket{\uparrow}\ket{\uparrow}$) decreases. The unequal
occupations of these triplet states lead thus to a nonequilibrium
spin accumulation in the dot, $P_{\ket{1,1}}-P_{\ket{1,-1}} < 0$,
which is shown in fig.~\ref{Fig:3}.
It is further interesting to note that in the antiparallel
configuration the possible non-spin-flip cotunneling processes are
proportional to $\Gamma_{\rm L}^{+}\Gamma_{\rm R}^{-}$ and
$\Gamma_{\rm L}^{-}\Gamma_{\rm R}^{+}$, whereas the spin-flip
cotunneling is proportional to $\Gamma_{\rm L}^{+}\Gamma_{\rm
R}^{+}$ and $\Gamma_{\rm L}^{-}\Gamma_{\rm R}^{-}$. It is clear
that the fastest cotunneling processes are the ones involving only
the majority spins, i.e. $\Gamma_{\rm L}^{+}\Gamma_{\rm R}^{+}$.
However, because of nonequilibrium spin accumulation, with
increasing the bias ($V>0$), the dot becomes dominantly occupied
by majority electrons of the right lead,
$P_{\ket{1,-1}}>P_{\ket{1,1}}$, and the processes proportional to
$\Gamma_{\rm L}^{+}\Gamma_{\rm R}^{+}$ are suppressed. As a
consequence, the differential conductance drops with the bias
voltage, leading to the zero-bias peak, see the dashed line in
fig.~\ref{Fig:2}(c). This is thus the nonequilibrium accumulation
of spin $S=1$ which is responsible for the maximum in $G$ at low
bias voltage. Because the dot is coupled symmetrically to the
leads, there is no spin accumulation in the parallel configuration
and the differential conductance exhibits a smooth parabolic
dependence on the bias voltage, see the solid line in
fig.~\ref{Fig:2}(c).

In the case of $\Delta_{\rm ST} >0$, at equilibrium the dot is
occupied by a triplet. Therefore, at low bias voltage,
$|eV|<|\Delta_{\rm ST}|$, transport is mediated by $S=1$ states.
When the voltage is increased, at $|eV| \approx |\Delta_{\rm
ST}|$, the occupation of the singlet state $\ket{0,0}_2$ increases
and all the four states start to participate in transport. This
leads a step in differential conductance which is clearly visible
in both magnetic configurations, see fig.~\ref{Fig:2}(d). The
width of the transport region where cotunneling through the
singlet state is suppressed is given by $2|\Delta_{\rm ST}|$. As
one can see in the figure, in the parallel configuration there is
one minimum, while in the antiparallel configuration there are two
minima separated by a zero-bias peak. The mechanism leading to the
peak at zero bias in the antiparallel configuration is the same as
in the aforementioned case of $\Delta_{\rm ST}=0$. Spin
accumulation induced in the dot, see fig.~\ref{Fig:3}, leads to a
decrease in differential conductance when increasing the bias
voltage from $V=0$.

The nonequilibrium spin accumulation in the antiparallel
configuration as a function of the bias voltage is displayed in
fig.~\ref{Fig:3}. In the case of $\Delta_{\rm ST}\geq 0$, spin
accumulation is always present for $V\neq 0$, while for
$\Delta_{\rm ST}<0$, spin accumulation appears when $|eV|\approx
|\Delta_{\rm ST}|$, i.e. at the onset of cotunneling through
triplet states. As pointed previously, spin accumulation is
responsible for the maximum in differential conductance. Thus, one
could expect that some maxima in $G$ should also appear in the
case of $\Delta_{\rm ST}<0$, where spin accumulation is induced
for $|eV|\geq|\Delta_{\rm ST}|$, see fig.~\ref{Fig:3}. This is
indeed visible in fig.~\ref{Fig:2}(a) and (b) where two small
maxima develop at $|eV|\approx |\Delta_{\rm ST}|$, symmetrically
with respect to the zero bias.

\begin{figure}[h]
  \center
  \includegraphics[width=0.42\columnwidth]{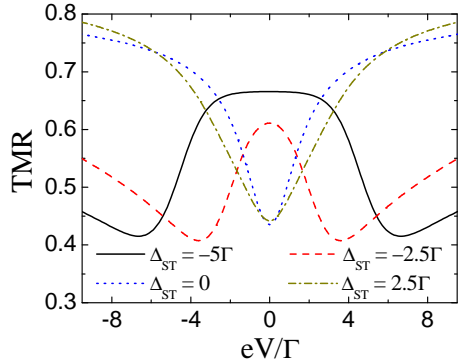}
  \caption{\label{Fig:4} (color online)
  Tunnel magnetoresistance
  as a function of the bias voltage for
  $\Delta_{\rm ST} = -5,-2.5,0,2.5 \Gamma$,
  as indicated in the figure.
  The other parameters are the same as in fig.~\ref{Fig:2}.}
\end{figure}

Another feature visible in fig.~\ref{Fig:2} is the difference
between conductance in the parallel and antiparallel magnetic
configuration. Conductance in the parallel configuration is
generally larger than that in the antiparallel configuration. This
is due to the asymmetry between the couplings of the dot levels to
the spin majority and spin minority electron bands in the
antiparallel configuration \cite{weymannPRB}. This difference
gives rise to tunnel magnetoresistance, ${\rm TMR} = (I_{\rm P} -
I_{\rm AP})/I_{\rm AP}$, where $I_{\rm P}$ ($I_{\rm AP}$) is the
current flowing through the system in the (anti)parallel
configuration \cite{weymannPRB,julliere75}. The bias voltage
dependence of TMR is shown in fig.~\ref{Fig:4}. As one can see,
TMR displays a nontrivial dependence on the ground state of the
dot. When the dot is occupied by a singlet, $\Delta_{\rm ST}<0$,
TMR at low bias displays a maximum plateau. With the same
assumptions as made when deriving eq.~(5), one finds that the
maximum in TMR for $\Delta_{\rm ST}<0$ and $|eV|<|\Delta_{\rm
ST}|$ can be approximated by, ${\rm TMR}^{S=0} = 2p^2/(1-p^2)$.
This value corresponds to the Julliere TMR, which is
characteristics of single tunnel junction \cite{julliere75}. In
our case, it results from the fact that in this transport regime
current flows due to non-spin-flip cotunneling. On the other hand,
for $\Delta_{\rm ST} \geq 0$, the TMR exhibits a minimum at zero
bias, which can be seen in fig.~\ref{Fig:4} for $\Delta_{\rm
ST}=0$ and $\Delta_{\rm ST} = 2.5\Gamma$.

\section{Summary}

We have considered spin-polarized transport through a doubly
occupied two-level quantum dot coupled to ferromagnetic leads in
the cotunneling regime. In this case, due to finite exchange
interaction, cotunneling current is mediated by singlet and
triplet states of the dot. We have shown that transport strongly
depends on the ground state of the system. If the dot is occupied
by a singlet at equilibrium, $G$ exhibits a minimum at low bias,
whose width is given by the strength of splitting between the
singlet and triplet states. In this transport regime the TMR is
given by the Julliere value, indicating that transport is
coherent. If triplet is the ground state, a zero-bias peak evolves
in differential conductance for antiparallel configuration. The
zero-bias maximum results from the nonequilibrium accumulation of
spin $S=1$ in the dot. The maximum is accompanied by a minimum in
TMR. The different behavior of differential conductance and TMR
may help in determining the ground state of the dot as well as the
splitting between the singlet and triplet states in the dot.

\acknowledgments

We acknowledge discussions with J\'ozef Barna\'s. The work was
supported by the Foundation for Polish Science.

%%%%%%%%%%%%%%%%%%%%%%%%%%%%%%%%%%%%%%%%%%%%%%%%%%%%%%%%%%%%%%%%
%%%%%%%%%%%%%%%%%%%%%%%%%%%%%%%%%%%%%%%%%%%%%%%%%%%%%%%%%%%%%%%%
%%%%%%%%%%%%%%%%%%%%%%%%%%%%%%%%%%%%%%%%%%%%%%%%%%%%%%%%%%%%%%%%


\begin{thebibliography}{0}

\bibitem{loss02}
  \Editor{D. D.~Awschalom, D.~Loss, \and N.~Samarth}
  \Book{Semiconductor Spintronics and Quantum Computation}
  \Publ{Springer, Berlin}
  \Year{2002}.

\bibitem{maekawa02}
  \Editor{S.~Maekawa \and T.~Shinjo}
  \Book{Spin Dependent Transport in Magnetic Nanostructures}
  \Publ{Taylor~\&~Francis}
  \Year{2002}.

\bibitem{kondo}
  \Name{R. L\'opez \and D. S\'anchez}
  \REVIEW{Phys. Rev. Lett.}{90}{2003}{116602};
  \Name{J. Martinek, Y. Utsumi, H. Imamura, J. Barna\'s, S. Maekawa, J. K\"onig, \and G. Sch\"on}
  \REVIEW{Phys. Rev. Lett.}{91}{2003}{127203};
  \Name{A. N. Pasupathy, R. C. Bialczak, J. Martinek, J. E. Grose, L.
  A. K. Donev, P. L. McEuen, \and D. C. Ralph}
  \REVIEW{Science}{306}{2004}{86}.

\bibitem{sequential}
  \Name{B. R. Bu\l ka}
  \REVIEW{Phys. Rev. B}{62}{2000}{1186};
  \Name{W. Rudzi\'nski \and J. Barna\'s}
  \REVIEW{Phys. Rev. B}{64}{2001}{085318};
  \Name{M. Braun, J. K\"onig, J. Martinek}
  \REVIEW{Phys. Rev. B}{70}{2004}{195345}.

\bibitem{cotunneling}
  \Name{D. V. Averin \and Yu. V. Nazarov}
  \REVIEW{Phys. Rev. Lett.}{65}{1990}{2446};
  \Name{K. Kang \and B. I. Min}
  \REVIEW{Phys. Rev. B}{55}{1997}{15412}.

\bibitem{weymanndong}
  \Name{I. Weymann, J. Barna\'s, J. K\"onig, J. Martinek, \and G. Sch\"on}
  \REVIEW{Phys. Rev. B}{72}{2005}{113301};
  \Name{B. Dong, X. L. Lei, Norman J. M. Horing} cond-mat/0509098.

\bibitem{weymannPRB}
  \Name{I. Weymann, J. K\"onig, J. Martinek, J. Barna\'s, \and G. Sch\"on}
  \REVIEW{Phys. Rev. B}{72}{2005}{115334};

\bibitem{twolevelQD}
  \Name{A. Thielmann, M. H. Hettler, J. K\"onig, \and G. Sch\"on}
  \REVIEW{Phys. Rev. B}{71}{2005}{045341};
  \Name{I. Weymann \and J. Barna\'s} cond-mat/0607039.

\bibitem{julliere75}
  \Name{M. Julliere}
  \REVIEW{Phys. Lett. A}{54}{1975}{225}.

\bibitem{FMSWCNT}
  \Name{J. R. Kim, H. M. So, J. J. Kim, \and J. Kim}
  \REVIEW{Phys. Rev. B}{66}{2002}{233401};
  \Name{A. Jensen, J. R. Hauptmann, J. Nyg\aa rd, \and P. E. Lindelof}
  \REVIEW{Phys. Rev. B}{72}{2005}{035419};
  \Name{S. Sahoo, T. Kontos, J. Furer, C. Hoffmann, M. Gr\"aber, A.
  Cottet, \and C. Sch\"onenberger}
  \REVIEW{Nat. Phys.}{1}{2005}{99};
  \Name{B. Nagabhirava, T. Bansal, G. U. Sumanasekera, \and B. W.
  Alphenaar, L. Liu}
  \REVIEW{Appl. Phys. Lett.}{88}{2006}{023503}.

\bibitem{SWCNT}
  \Name{D. H. Cobden \and J. Nyg\aa rd}
  \REVIEW{Phys. Rev. Lett.}{89}{2002}{046803};
  \Name{W. Liang, M. Bockrath, \and H. Park}
  \REVIEW{Phys. Rev. Lett.}{88}{2002}{126801}.

\bibitem{hamiltonian}
  \Name{W. Izumida, O. Sakai, \and S. Tarucha}
  \REVIEW{Phys. Rev. Lett.}{87}{2001}{216803};
  \Name{B. Dong \and X. L. Lei}
  \REVIEW{Phys. Rev. B}{66}{2002}{113310}.

\bibitem{kogan04}
  \Name{A. Kogan, S. Amasha, D. Goldhaber-Gordon, G. Granger, M. A.
  Kastner, \and H. Shtrikman}
  \REVIEW{Phys. Rev. Lett.}{93}{2004}{166602}.


\end{thebibliography}
\end{document}